# DUAL-FREQUENCY RADAR WAVE-INVERSION FOR SUB-SURFACE MATERIAL CHARACTERIZATION

*Ishfaq Aziz[1], Elahe Soltanaghai[2], Adam Watts[3], Mohamad Alipour[1]*

[1]Civil and Environmental Engineering, University of Illinois Urbana Champaign, [2]Computer Science, University of Illinois Urbana Champaign, [3]Pacific Wildland Fire Sciences Laboratory, United States Forest Service

**ABSTRACT**

Moisture estimation of sub-surface soil and the overlaying biomass layer is pivotal in precision agriculture and wildfire risk assessment. However, the characterization of layered material is nontrivial due to the radar penetration-resolution tradeoff. Here, a waveform inversion-based method was proposed for predicting the dielectric permittivity (as a moisture proxy) of the bottom soil layer and the top biomass layer from radar signals. Specifically, the use of a combination of a higher and a lower frequency radar compared to a single frequency in predicting the permittivity of both the soil and the overlaying layer was investigated in this study. The results show that each layer was best characterized via one of the frequencies. However, for the simultaneous prediction of both layers' permittivity, the most consistent results were achieved by inversion of data from a combination of both frequencies, showing better correlation with *in situ* permittivity and reduced prediction errors.

## 1. INTRODUCTION

Precision agriculture, effective wildfire risk assessment, subsurface condition monitoring, and infrastructure planning depend on the quality and accuracy of estimating sub-surface properties like soil moisture content, biomass properties, and depth of sub-surface layers. Natural soil is often covered by an overlaying top layer of organic or woody material. The estimation of moisture levels of both the soil layer and the top organic layer is crucial for identifying the occurrence and extent of wildfires [1]. The upper organic layer serves as fuel for wildfires, and measurements of surface fuel are fundamental inputs for the development of fuel mapping systems [2], [3], [4], [5], [6], [7]. Hence, the estimation of properties of both the soil layer and the overlaying top layer carries significant importance.

Traditional methods of measuring soil moisture involve performing gravimetric tests that are time-consuming and costly or using *in situ* sensors like the time-domain reflectometry (TDR) sensors that are hard to deploy. Moreover, direct determination of soil moisture through these methods becomes challenging in cases where the top organic layer is present above the soil. Non-intrusive measurements of sub-surface properties through active and passive remote sensing have been proposed to address these problems. This study employs active radar sensing to estimate the moisture properties of the two layers: the soil layer and the overlaying layer. Primarily, the dielectric permittivity of each layer is estimated from radar signals, which is highly correlated with the moisture contents of the layers [8], [9], [10], [11]. Commonly used techniques for estimating permittivity, moisture, and other parameters from radar signals include data-driven methods and waveform inversion [12], [13], [14], [15], [16]. A model-updating-based waveform inversion method was proposed in this study to estimate the layer permittivities from radar data. The unknown parameters were estimated by iteratively updating a numerical model while pushing its divergence from experimental results to zero via an optimization scheme. In this study, Bayesian optimization was employed to estimate permittivity values from radar waveforms.

Estimating permittivity (or other parameters) from a radar scan depends on the quality, depth, and resolution of data collected by a radar, which in turn depends on the radar frequency. To accurately predict the permittivity of a single soil layer, selecting one appropriate frequency based on the expected depth of the layer shall be sufficient. However, predicting the permittivity of multiple layers with only one radar frequency can introduce potential challenges. For instance, using a high frequency would limit penetration depth, resulting in less information from the bottom layers. Similarly, using a lower frequency would produce a signal with a lower axial resolution. This is depicted in Figure 1, which shows an amplitude envelope of a received radar signal for a one-layer material (Figure 1: a and b) and signals received by radars of two different frequencies (Figure 1: c and d) for varying depth and permittivity of the material layer. Figure 1(c) shows that for low depth and low permittivity, the direct wave and the reflected wave from the bottom are not clearly separated due to the low resolution of the lower frequency (0.7 GHz) signal. However, for a high-frequency (2 GHz) radar (Figure 1d), these two amplitude peaks are well separated, indicating higher resolution.

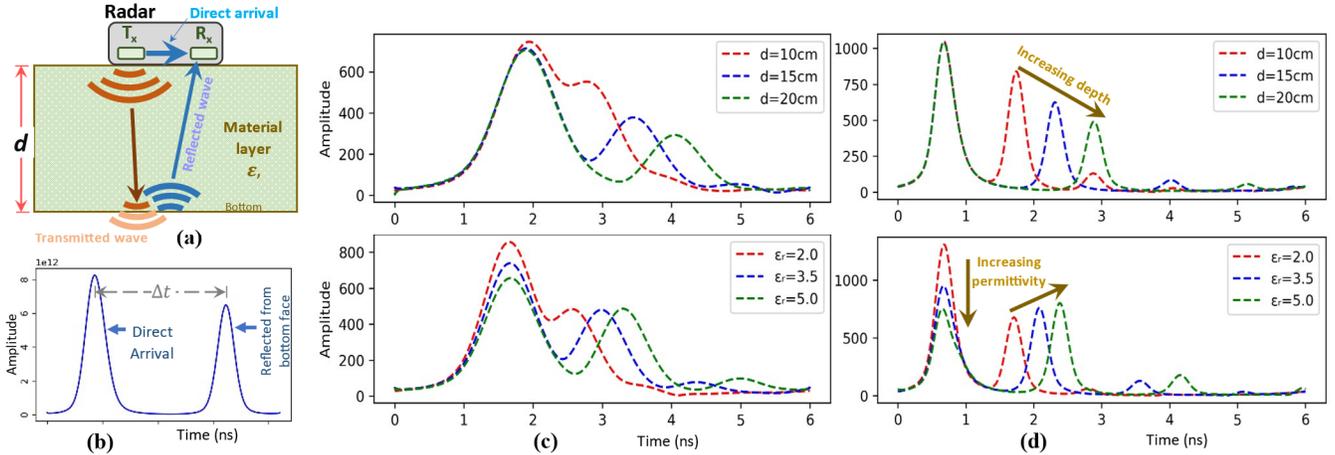

Figure 1. (a) Radar scan, (b) Amplitude envelope, (c) Received signals of **0.7 GHz** radar due to variation in depth: *d* and permittivity: $\varepsilon_r$, (d) Received signals of **2.0 GHz** radar due to variation in depth: *d* and permittivity: $\varepsilon_r$.

Investigations using a high-frequency radar (2.7GHz) showed that, at high moisture levels and for greater depths (>25 cm including soil and top layer), the radar signal is attenuated, indicating weak reflection from the bottom, resulting in inconsistent prediction of sub-surface parameters [17]. In this case, as the penetration capability of low-frequency radar is higher, it is helpful in investigating deep layers. Hence, different frequencies offer different sensing capabilities.

In essence, while a lower frequency signal can be useful for a deeper layer, a higher frequency can provide high-resolution data from the shallower layer. Hence, to improve the prediction of the permittivity of both layers for a wide range of layer depth and soil permittivity, the capabilities of both high and low-frequency radars need to be leveraged concurrently. With this view, waveform inversion of a combination of low and high-frequency radar scans was investigated in this study to accurately predict the permittivity of two layers, which can be converted to moisture contents using empirical equations like Topp's equations [11]. The results of predictions from dual-frequency radar waveform inversion were compared to predictions obtained when only a single-frequency radar scan was used.

## 2. METHODOLOGY

This paper utilizes a model-updating-based waveform-inversion approach to determine the permittivity of sub-surface layers from radar signals. The schematic of the model updating is shown in Figure 2. First, experimental data was collected with radar from the top of sub-surface layers consisting of soil and an overlaying layer of organic/woody materials. Then, a numerical model was created using the finite difference time domain (FDTD) method. The initial radar response received in the numerical simulation does not match the experimental signal due to the use of random initial parameters. The numerical model was then updated iteratively using Bayesian optimization to match the observed signal. The permittivity values of the sub-surface layers were changed at each iteration to update the numerical model and reduce an objective function encompassing the discrepancy between the model and experiment. The objective function used in the optimization algorithm is the relative error ($RE$) (Equation 1), which is the relative difference between the experimental response ($y$) and the numerical response ($\bar{y}$). The dielectric permittivity of both the soil layer and the overlaying organic layer was estimated by model-updating.

$$RE = \sqrt{\frac{\sum(y-\bar{y})^2}{\sum y^2}} \times 100\% \qquad \text{Equation 1}$$

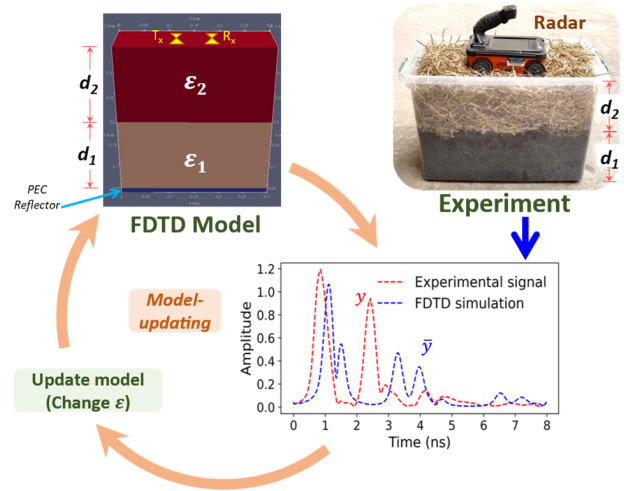

Figure 2. Schematic of proposed model-updating.

In this study, parameter estimation was conducted for three different cases. The first case was to apply the proposed method to data collected using only a high-frequency radar. The second case was similar to the first case, except that a low-frequency radar was used. The third case included the combination of both low and high-frequency radar for parameter estimation. The predictions from all cases were compared with permittivity measured by TDR.

## 3. EXPERIMENTAL INVESTIGATION

In a previous study conducted by the authors [17], the proposed model updating technique was shown to produce good agreement with TDR-measured values up to an overlaying layer depth of 10 cm. However, if the depth of the overlaying layer increases, the predictions deteriorate. Hence, dual frequency radar scans are conducted in the present study to investigate the performance of the proposed technique in the presence of deeper material layers (20 cm soil and 20 cm top layer).

The first step of the experimental investigation was to calibrate the radars for their accurate modeling in high-fidelity FDTD simulations. The two radars used in this study were manufactured by GSSI and were listed to have center frequencies of 900 and 2700 MHz, respectively. However, the frequency bands of both radars are quite broad, for which the center frequency may vary from the specified values. In addition, the types of waveforms transmitted by these radars are usually unknown. Hence, to find the frequency and waveform types that represent the radar signals in FDTD simulation, optimization of the antenna pulse parameters was conducted via FDTD simulations. The process of antenna pulse optimization is elaborately shown in [17]. The optimum center frequencies of the higher and the lower frequency radar were found to be 1.578 GHz and 0.7 GHz in FDTD simulation, respectively. The optimum waveform type for both radars was Gaussian. The remainder of the numerical simulations in this study were carried out with the retrieved optimum pulse parameters.

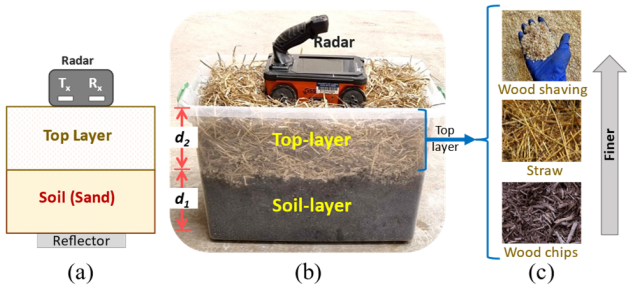

Figure 3. (a) Schematic view of experimental investigation, (b) Experimental setup, (c) Top layer variation

Next, to evaluate the proposed method, experimental investigations were carried out for a diverse range of sub-surface parameters. The experimental setup is shown in Figure 3, where the bottom 20 cm deep soil layer is overlaid by a top 20 cm deep woody/organic layer. In this investigation, six different soil moisture contents and three different top-layer materials with varying levels of fineness were used. The three top layer materials are wood chips, straw, and wood shavings, having mean particle sizes of 11.7 mm, 6.4 mm, and 2.5 mm, respectively. Hence, the experimental matrix was composed of 3 top layers and 6 moisture contents, thus creating 18 radar scan cases.

## 4. RESULTS

The prediction of the permittivity of soil and the top layer is shown in Figure 4 and Figure 5, respectively.

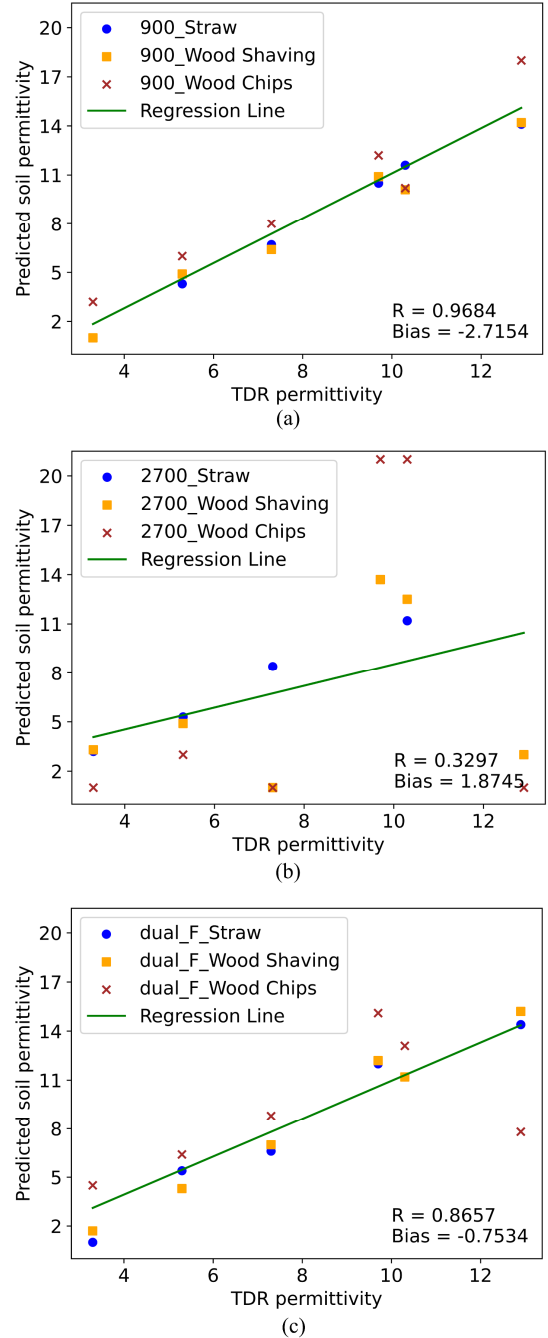

Figure 4. Prediction of soil permittivity in the presence of 20 cm top layer. (a) With only 900 MHz antenna, (b) With only 2700 MHz antenna, (c) With combination of both antenna

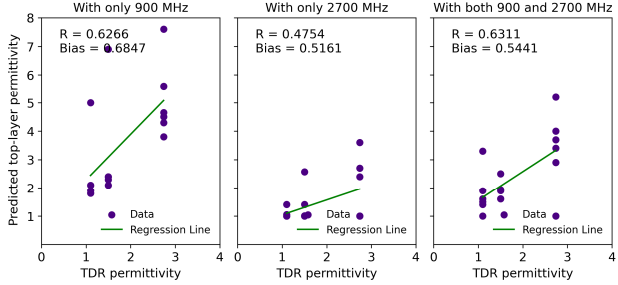

Figure 5. Prediction of top-layer permittivity.

Table 1. Errors in predicting top layer permittivity.

| Error Metric | Using only 900 MHz | Using only 2700 MHz | Using both frequencies |
|---|---|---|---|
| MRE | 1.05 | 0.29 | 0.45 |
| MAE | 1.76 | 0.58 | 0.78 |
| RMSE | 2.28 | 0.82 | 1.04 |

The correlations of predicted permittivity values with values measured by the TDR sensor are also shown in the same graphs. For the case of predicting soil permittivity, the correlation ($R$-value) is highest when prediction is done using only the low-frequency radar and lowest when prediction was done with only the high-frequency radar. Acceptable correlation is observed when prediction was done using a combination of data from both high and low-frequency radar (labeled as 'dual-F' in Figure 4).

The prediction of top-layer permittivity (Figure 5) shows that the best correlation was achieved when a combination of frequencies was used. However, as shown in Table 1, the mean error metrics (mean relative error, MRE; mean absolute error, MAE; and root mean squared error, RMSE) in predicting top-layer permittivity were the lowest for the case of using only the high-frequency radar and the low-frequency radar produced the highest errors. The combination of frequencies produced better error than the case of using 900 MHz only.

## 5. DISCUSSIONS

The correlation between the predicted and measured bottom soil layer permittivity is shown to be the highest when using only the low-frequency radar and lowest when using the high-frequency radar. This is because the total depth of the layers (20 cm soil + 20 cm top layer) is high compared to the penetration capability of the high-frequency radar. In addition, at higher moisture content, the signals get even more attenuated, and thus, predictions get deteriorated [17]. However, the simultaneous use of both frequencies results in an acceptable compromise in terms of the correlation between predicted and measured permittivity.

The prediction of the top layer permittivity was more consistent with TDR permittivity when a combination of frequencies was used, while the errors in predicting the top layer permittivity were lowest for the case of using only the higher frequency. This is because the resolution of this radar is higher, which enables the distinction of reflection from the bottom of the top layer from the direct wave of radar that travels from transmitter to receiver.

Overall, to simultaneously predict the permittivity of both layers, the combination of low and high frequencies produced satisfactory prediction in terms of both correlation with TDR permittivity and mean absolute error. While the prediction of deeper layer parameters was better with a lower frequency, the higher frequency was more effective for the top layer characterization. This is consistent with the widely known tradeoff between high resolution and deeper penetration encountered while choosing a radar frequency. This study, therefore, highlights the importance of the frequency specifications of the system for effective multi-layer subsurface sensing applications.

## 6. CONCLUSIONS

A model-updating-based approach was proposed to estimate sub-surface parameters, including soil permittivity in the presence of a top organic layer. The permittivity of the top layer was also predicted with the proposed method. This study investigated the feasibility of using a combination of frequencies to estimate sub-surface parameters through the model-updating-based approach. Due to the higher depth of material layers, the permittivity of bottom soil layers was better predicted by the lower frequency radar, while a combination of both higher and lower frequencies was more effective for top layer characterization. The error metrics in predicting absolute permittivities of the top layer were, however, lowest when prediction was done using the higher frequency. To characterize both layers simultaneously, using the combination of high and low frequency performed the best in terms of agreement with TDR-measured values and errors. Hence, for simultaneous estimation of parameters of multiple layers of varying depth, the combination of frequencies is expected to be a preferable choice than using single-frequency radar data.

Further studies should be conducted to use multi-frequency waveform inversion to estimate other sub-surface parameters, such as the depth of the top layer, and electrical conductivities of each layer. The accurate parameter prediction of both soil layers and the top layer will play an important role in applications spanning precision agriculture, moisture mapping, wildfire assessment, and sub-surface monitoring.

## 7. ACKNOWLEDGEMENT

The authors acknowledge partial funding for this study by the U.S. Forest Service and Keysight Technologies. The authors also acknowledge NVIDIA's Academic Hardware Grants Program for GPU support, technical support by Mr. Matthew Ware, Mr. Matthew Smith, and the GSSI technical team.